\documentclass[floatfix,aps,superscriptaddress,twocolumn,prb]{revtex4-1}  
\usepackage{bm}
\usepackage{graphicx}
\usepackage[usenames]{color}
\usepackage{microtype}
\usepackage{amsmath}
\usepackage{amssymb}
\usepackage{xspace}
\usepackage{footnote}
\usepackage{relsize}
\usepackage{hhline}
\usepackage{mhchem}

\usepackage{physics}
\usepackage{comment}
\usepackage[normalem]{ulem}

\newcommand{\mbf}[1]{\ensuremath{\mathbf{#1}}}

\newcommand{\br}{\mbf{r}}
\newcommand{\bx}{\mbf{x}}

\newcommand{\D}[2][]{\ensuremath{\mathop{}\!\mathrm{d}^{#1}{#2}\,}}

\newcommand{\Rhat}{\hat{R}}
\newcommand{\Othree}{{O(3)}}
\newcommand{\SOthree}{{SO(3)}}

\newcommand{\bC}{\mbf{C}}
\newcommand{\bU}{\mbf{U}}

\newcommand{\bv}{\mbf{v}}

\begin{document}

\title{Recursive evaluation and iterative contraction of $N$-body equivariant features}

\author{Jigyasa Nigam}
\thanks{These authors contributed equally to this work}
\affiliation{Laboratory of Computational Science and Modeling, IMX, \'Ecole Polytechnique F\'ed\'erale de Lausanne, 1015 Lausanne, Switzerland}
\affiliation{National Centre for Computational Design and Discovery of Novel Materials (MARVEL), {\'E}cole Polytechnique F{\'e}d{\'e}rale de Lausanne, 1015 Lausanne, Switzerland}

\author{Sergey Pozdnyakov}
\thanks{These authors contributed equally to this work}
\affiliation{Laboratory of Computational Science and Modeling, IMX, \'Ecole Polytechnique F\'ed\'erale de Lausanne, 1015 Lausanne, Switzerland}

\author{Michele Ceriotti}
\email{michele.ceriotti@epfl.ch}
\affiliation{Laboratory of Computational Science and Modeling, IMX, \'Ecole Polytechnique F\'ed\'erale de Lausanne, 1015 Lausanne, Switzerland}
\affiliation{National Centre for Computational Design and Discovery of Novel Materials (MARVEL), {\'E}cole Polytechnique F{\'e}d{\'e}rale de Lausanne, 1015 Lausanne, Switzerland}

\date{\today}
\begin{abstract}
Mapping an atomistic configuration to an $N$-point correlation of a field associated with the atomic positions (e.g. an atomic density) has emerged as an elegant and effective solution to represent structures as the input of machine-learning algorithms. 
While it has become clear that low-order density correlations do not provide a complete representation of an atomic environment, the exponential increase in the number of possible $N$-body invariants makes it difficult to design a concise and effective representation.
We discuss how to exploit recursion relations between equivariant features of different orders (generalizations of $N$-body invariants that provide a complete representation of the symmetries of improper rotations) to compute high-order terms efficiently. In combination with the automatic selection of the most expressive combination of features at each order, this approach provides a conceptual and practical framework to generate systematically-improvable, symmetry adapted representations for atomistic machine learning. 
\end{abstract}

\maketitle

\section{Introduction}
Equivariant, atom-centred structural representations have driven the progress of atomistic machine-learning over the last decade~\cite{behl-parr07prl,bart+10prl,bart+13prb,thom+15jcp,de+16pccp,shap16mms,glie+18prb,fabe+18jcp,zhan+18prl,will+19jcp,drau19prb, ferre2017graph-kernels}. These representations preserve the transformation rules of the target property with respect to fundamental symmetries such as translation, rotation, inversion and permutation of identical atoms.
Such atomic descriptions have found widespread usage because the incorporation of symmetries (as well as the locality that derives from their atom-centred nature) make the data-driven regression models more transferable and efficient when learning invariant (or covariant~\cite{glie+17prb,gris+18prl}) target properties. 
Most of these equivariant representations can be seen as projections of the many-body correlation functions of a decorated atom density onto more or less arbitrary choices of bases, and linear regression models based on these features are equivalent to a body-ordered expansion of the target property~\cite{glie+18prb,will+19jcp,drau19prb,jinn+20jcp}.
The expansion is typically truncated at the third or fourth order-correlation~\cite{bart+13prb,thom+15jcp}, which is problematic because low-order correlations are incomplete~\cite{pozdnyakov2020completeness}, so that one can build configurations which have the same features despite having different structure and properties.
With increasing body order, however, the number of terms in the projection grows exponentially.

In this Communication we discuss a recursive construction for equivariant features, that avoids some of the formally (and computationally) daunting expressions that one encounters when writing explicitly the form of high-order invariant features~\cite{singer2006construction,drau19prb,drau20preprint}, while simplifying a discussion of the relations between different body orders. To prevent the exponential increase in the total number of features, we then introduce an $N$-body iterative contraction of equivariants (NICE) framework, and demonstrate it on a simple -- yet challenging -- benchmark dataset.

\section{Theory} 

The notation we use is a refinement of that introduced in Refs.~\citenum{will+18pccp,will+19jcp}. The braket $\bra*{I}\ket*{A}$ indicates a feature (labelled by $I$) which is meant to describe a structure and its associated properties (labelled by $A$). Both indices can be expressed in a contracted or expanded form, depending on the level of detail that is needed for a given manipulation.
We start by defining $(\nu+1)$-body equivariants as averages of the atom-centred density over the $\SOthree$ group
\begin{multline}
\bra*{n_1 l_1 m_1; \ldots n_\nu l_\nu m_\nu; LM}\ket*{\rho_i^{\otimes \nu} \lambda \mu}_\SOthree = \\[-1mm]
\int \D{\Rhat} 
\prod_j\bra*{n_j l_j m_j}\Rhat\ket*{\rho_i} \bra*{LM}\Rhat\ket*{\lambda\mu}=\delta_{L\lambda}\times\\[-3mm]
\sum_{m_1'\ldots m_\nu'}
\int\D{\Rhat}
\prod_j\bra*{n_j l_j m_j'}\ket*{\rho_i}
D^{l_j}_{m_j m'_j}(\Rhat) D^\lambda_{M\mu}(\Rhat),
\label{eq:haar-definition}
\end{multline}
where $\nu$ indicates the number of densities included in the correlation function, and $D^l_{mm'}(\Rhat)$ is the Wigner matrix associated with the rotation $\Rhat$. The ket $\ket*{\rho_i^{\otimes \nu} \lambda \mu}_\SOthree$ indicates that for each feature $\bra*{I}$ we must compute a set of $2\lambda +1 $ entries, labelled by $\mu$, that transform under rotations as the spherical harmonic of order $\lambda$~\cite{gris+18prl}.
The term $\bra*{n l m}\ket*{\rho_i}$ indicates an expansion in radial functions and spherical harmonics of the atom density centred on atom $i$,
\begin{equation}
\bra*{n l m}\ket*{\rho_i}=
\int \D{\bx} R_n(x) Y^l_m(\hat{\bx}) 
\sum_{j} g(\bx - \br_{ji}) 
\end{equation}
where $g$ is a localized function (possibly a Dirac $\delta$, a limit for which this formulation reduces to the atomic cluster expansion~\cite{drau19prb}) and $\br_{ji}=\br_j-\br_i$ is the distance vector between atoms $i$ and $j$. 
This construction is easily extended to multiple atomic species, as well as to other attributes of the atoms, by computing a tensor product between the atom density and these additional quantities. The indices associated with the atomic nature must be gathered together with the radial index $n$ - so all of the developments in this work apply equally well to the more general case by considering $n$ as a compound index.
As discussed in the SI, Eq.~\eqref{eq:haar-definition} is somewhat redundant: $L$ is bound to be equal to $\lambda$, $M$ can be fixed to any value and only introduces some inconsequential phases, and there are constraints on the values of the $m_i$ - corresponding to the loss of degrees of freedom associated with the covariant integration. For instance, $\sum_j m_j=-M$, and $\sum_j m_j' = -\mu$.

\subsection{Recursive construction of equivariant features}

To obtain a more transparent and concise enumeration of the $N$-body equivariants, we devise a labeling that makes it simpler to identify those that are linearly independent, and a recursion relation to build them efficiently. To fully describe the symmetries of each equivariant, we also introduce a label $\sigma$ that indicates their parity with respect to inversion\footnote{$\Othree$ equivariants can be obtained symmetrizing over both $\Rhat$ and $\hat{i}$, a field built as the tensor product of $\nu$ atom-centred densities, a spherical harmonic, and a parity field $\ket{\sigma}$ that is invariant under rotations and behaves as a scalar/tensor ($\sigma=1$) or a pseudoscalar/pseudotensor ($\sigma=-1$) under inversion.}.

We start defining the $\nu=1$ equivariants as,
\begin{multline}
\bra*{n_1 l_1 k_1}\ket*{\rho_i^{\otimes 1} \lambda\mu\sigma}_\Othree \equiv
\bra*{n_1 \lambda \,{(-\mu)}}\ket*{\rho_i}  
\delta_{l_1\lambda} \delta_{k_1 \lambda}
\delta_{\sigma 1}.
\label{eq:rho-one}
\end{multline}
Given that many indices are redundant and that all $\nu=1$ terms behave with $\sigma=1$ parity, we also introduce the shorthand notation $\bra*{n_1}\ket*{\rho_i^{\otimes 1}\lambda\mu}\equiv\bra*{n_1 l_1 k_1}\ket*{\rho_i^{\otimes 1} \lambda\mu\sigma}_\Othree$
Higher order terms can be obtained using an iterative formula modeled after the addition of angular momenta
\begin{multline}
\bra*{\ldots; n_\nu l_\nu k_\nu; n l k}\ket*{\rho_i^{\otimes (\nu+1)} \lambda \mu \sigma}_\Othree = \delta_{\sigma((-1)^{l+k+\lambda}s)} c_{k\lambda}  \times \\
\sum_{qm}  \bra*{l m; k q}\ket*{\lambda\mu} \bra*{n}\ket*{\rho_i^{\otimes 1} l m} 
\bra*{\ldots; n_\nu l_\nu k_\nu}\ket*{\rho_i^{\otimes \nu} k q s},
\label{eq:rho-recursion}
\end{multline}
where $ \bra*{l m; k q}\ket*{\lambda\mu}$ is a Clebsch-Gordan coefficient and the scaling factor $c_{ll'} = \sqrt{(2l+1)/(2l'+1)}$.
The index $\sigma$ tracks the parity of the equivariants, ensuring that terms with an even $\lambda+\sum_j l_j$ are associated with $\sigma=+1$ and those with an odd sum with $\sigma=-1$. The mapping between the  integral form~\eqref{eq:haar-definition} and the recursive form~\eqref{eq:rho-recursion} is not entirely trivial, and is derived in the SI.\footnote{Given that $k_1$ and $k_2$ cannot be varied independently from $l_1$ and $l_2$, we suggest to use a compact notation $\bra*{n_1l_1; n_2 l_2; n_3 l_3 k_3 \ldots}$, similar to the one for the $\nu=1$ term in Eq.~\eqref{eq:rho-one}.}

\paragraph*{Linearly independent covariants}
This construction makes it easy to determine which terms are linearly independent: well-established angular-momentum theory results~\cite{topic5-recouplingtheory} show that one only needs to consider terms where the $l$ indices are sorted in ascending order; if the density is expanded up to an angular momentum cutoff $l_\text{max}$, this implies $l_1\le l_2\le l_3...\le l_\nu\le l_\text{max}$.
When two $l$ indices are equal, the covariants are symmetric with respect to an exchange of the corresponding $n$ indices. Thus, although in general the $n$ indices need not be sorted, whenever $l_\nu=l_{\nu+1}$ the terms with $n_{\nu+1}<n_\nu$ can be discarded.

\paragraph*{Polynomially-independent invariants.}
The case of \emph{invariant} features is particularly important, as they provide a basis to expand properties such as the potential energy that are left unchanged by rigid rotations and inversion. Eq.~\eqref{eq:rho-recursion} shows clearly how they can also be obtained efficiently by keeping track of all the equivariants of lower-order to compute $\ket*{\rho_i^{\otimes (\nu+1)}}\equiv \ket*{\rho_i^{\otimes (\nu+1)}001}$
\begin{multline}
\bra*{\ldots n_\nu l_\nu k_\nu; n l}\ket*{\rho_i^{\otimes (\nu+1)}} = 
\sum_m \bra*{l {m}; l (-m)}\ket*{00} \times\\[-2mm]
\bra*{n}\ket*{\rho_i^{\otimes 1} l m} \bra*{\ldots n_\nu l_\nu k_\nu}\ket*{\rho_i^{\otimes \nu} l {(-m)} 1}
,
\label{eq:rho00-recursion}
\end{multline}
an expression that encompasses neatly the well-known formulas for the SOAP power spectrum and bispectrum~\cite{bart+13prb}. 
In the case of invariant features, in addition to the rules that identify linearly independent terms based on angular-momentum theory, it is also relevant whether higher-order terms can be written as polynomials of lower-order invariants. This is because  non-linear regression schemes (kernel methods, polynomial regression or neural networks) produce arbitrary polynomial combinations of the low-order invariants. Thus, terms that \emph{cannot} be written as polynomials of lower invariants are likely to be more informative, and more worthy of being retained for use in non-linear regression. 
When one of the intermediate couplings $k$ is zero, the higher-body order term becomes a polynomial of two lower-body terms as, 
\begin{multline}
\braket*{n_1 l_1;\ldots n_p l_p k_p=0; \ldots; n_\nu l_\nu}{\rho_i^{\otimes\nu}} = \\
 \braket*{n_1 l_1;\ldots n_{p-1} l_{p-1}}{\rho_i^{\otimes(p-1)}}  \braket*{n_p l_p; \ldots n_\nu l_\nu}{\rho_i^{\otimes(\nu-p+1)}} 
\end{multline}
Furthermore, if at least two l's of the spherical harmonic basis of expansion are chosen to be the same, say $l_p=l_{p+1}=l$, then the elements of the set of projections, $\braket*{n_1l_1; \ldots n_p l_p k_p; n_{p+1} l_{p} k_{p+1}; \ldots; n_\nu, l_\nu}{\rho_i^{\otimes\nu}} $ are not linearly independent, as discussed in the SI.

\subsection{Iterative construction of contracted equivariants}

\begin{figure}
    \centering
\includegraphics[width=1.0\linewidth]{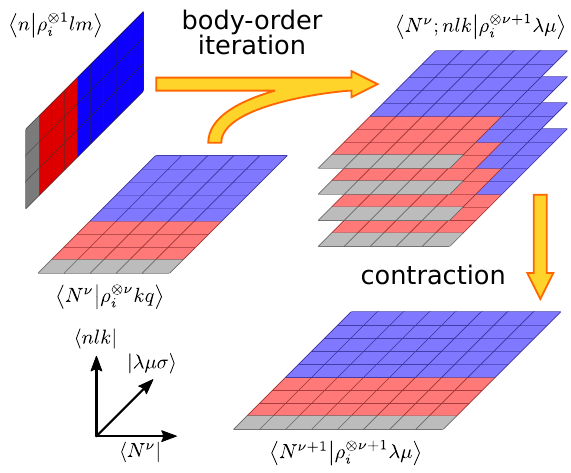}
\caption{A schematic representation of the NICE framework. A hierarchy of $N$-body equivariant features is built by iterative combination with the atom density coefficients, and the exponential increase in feature space size is kept at bay by successive contractions.}
    \label{fig:nice-scheme}
\end{figure}

Even though Eq.~\eqref{eq:rho-recursion} makes it possible to compute a given equivariant feature with a cost that scales only linearly with the body order, and even though linear and polynomial relationships between features allow one to discard several terms, the number of independent features scales exponentially with $\nu$, making it impractical to enumerate all the equivariants. Selecting the most important terms for a given application is therefore crucial to obtain a viable scheme based on high-order features.
Several of such schemes have been proposed~\cite{imba+18jcp,wood-thom18jcp,thom+15jcp,seko2014sparse, li2018-lassoselection,gastegger2018wacsf,chen2017accurate}, either based on selecting the most relevant features from a large pool of candidates (that still requires computing all of them at least in a preliminary phase), by selecting a subset based on heuristic arguments, or by designing a neural network architecture that incorporates $\SOthree$ combination rules analogous to~\eqref{eq:rho-recursion}\cite{anderson2019cormorant, thomas2018tensor, kondor2018n}.
We propose a strategy to generate a set of high-order equivariants based on a combination of an iteration rule and a simple selection scheme based on principal component analysis, that serves as a proof of principle for more sophisticated schemes based on linear or non-linear feature extraction (Fig.~\ref{fig:nice-scheme}).

\paragraph*{Feature selection/contraction.}
Assume that a pool of $\Othree$ equivariant features has been computed for a given order $\nu$. 
We disregard the internal structure of the featurization, and simply tag these features as $\bra*{N}\ket*{\rho_i^{\otimes\nu}\lambda\mu\sigma}$.
We look for a contraction that extracts the largest amount of (linearly) independent information. The most straightforward approach is to compute correlation matrices
\begin{equation}
C^{\nu; \lambda\sigma}_{NN'} = \sum_{Ai\mu} 
\bra*{N}\ket*{A;\rho_i^{\otimes\nu}\lambda\mu\sigma}\bra*{A;\rho_i^{\otimes\nu}\lambda\mu\sigma}\ket*{N'},
\label{eq:covariance}
\end{equation}
where the sum runs over all structures $A$ and environments $i$ in a reference data set.
This matrix can then be diagonalized as $\bC^{\nu;\lambda\sigma}=\bU\operatorname{diag}(\bv)\bU^T$, and the most significant features built as 
\begin{equation}
\bra*{\tilde{N}^{\nu;\lambda\sigma}}\ket*{\rho_i^{\otimes\nu}\lambda\mu\sigma} = \sum_N U^{\nu;\lambda\sigma}_{N\tilde{N}} \bra*{N}\ket*{\rho_i^{\otimes\nu}\lambda\mu\sigma}.
\label{eq:contraction}
\end{equation}
For clarity of exposition we consider a correlation matrix built exclusively on terms of order $\nu$, but it would clearly be possible to combine all of the features that have been retained up to the current body order iteration. %

\paragraph*{Body order iteration.}
The contracted features can be combined with the density coefficients to build a set of $(\nu+1)$-order equivariants, by straightforward application of Eq.~\eqref{eq:rho-recursion}:
\begin{multline}
\bra*{\tilde{N}^{\nu;ks}; n l k}\ket*{\rho_i^{\otimes (\nu+1)} \lambda \mu \sigma}_\Othree = \delta_{\sigma((-1)^{l+k+\lambda}s)} c_{k\lambda}  \times \\
\sum_{qm} \bra*{l m; k q}\ket*{\lambda\mu} \bra*{n}\ket*{\rho_i^{\otimes 1} l m} 
\bra*{\tilde{N}^{\nu;ks}}\ket*{\rho_i^{\otimes \nu} k q s}.
\label{eq:contract-recursion}
\end{multline}
Pooling together all of the feature indices when constructing the correlation matrix mixes the $nlk$ channels, which makes it impossible to keep track of the trivial linear dependencies between angular momentum combinations. 
They are however identified automatically by the contraction step, together with non-trivial correlations between the features, and are immediately discarded.
Another important consideration is that, for a given angular cutoff of the density expansion, each application of~\eqref{eq:contract-recursion} doubles the maximum possible value of $\lambda$. To prevent an exponential increase in the number of covariant terms, we cutoff $\lambda$ to the same $l_\text{max}$ used for the density expansion. 

We also want to stress that this scheme is just one of the many conceivable combinations of a body-order recursion and feature-selection steps. Disregarding completely the physical significance of the $nlk$ indices, one can generate high-order features by combining lower-order equivariants using the sum rules for angular momenta, similar to what is done in covariant neural networks~\cite{anderson2019cormorant}.
One can also introduce, at each level, an arbitrary non-linear function of the \emph{invariant} terms of lower order, as suggested in Ref.~\citenum{will+19jcp}, yielding an expression of the form 
\begin{multline}
\bra*{N N' lk}\ket*{f \rho_i^{\otimes (\nu+\nu')}\lambda \mu} = 
f\left[\left\{ \bra*{N}\ket*{\rho_i^{\otimes\nu}},\bra*{N'}\ket*{\rho_i^{\otimes\nu'}}\right\}\right]\times\\
\sum_{mq} \bra*{lm; kq}\ket*{\lambda\mu}
\bra*{N}\ket*{\rho_i^{\otimes\nu}lm}\bra*{N'}\ket*{\rho_i^{\otimes\nu'}kq}.
\label{eq:j-sum}
\end{multline}
We name the family of schemes that builds body-order equivariants using a sequence of angular-momentum iterations and feature selection the $N$-body iterative contraction of equivariants (NICE) framework.

\begin{figure}
    \centering
    \includegraphics[width=1.0\linewidth]{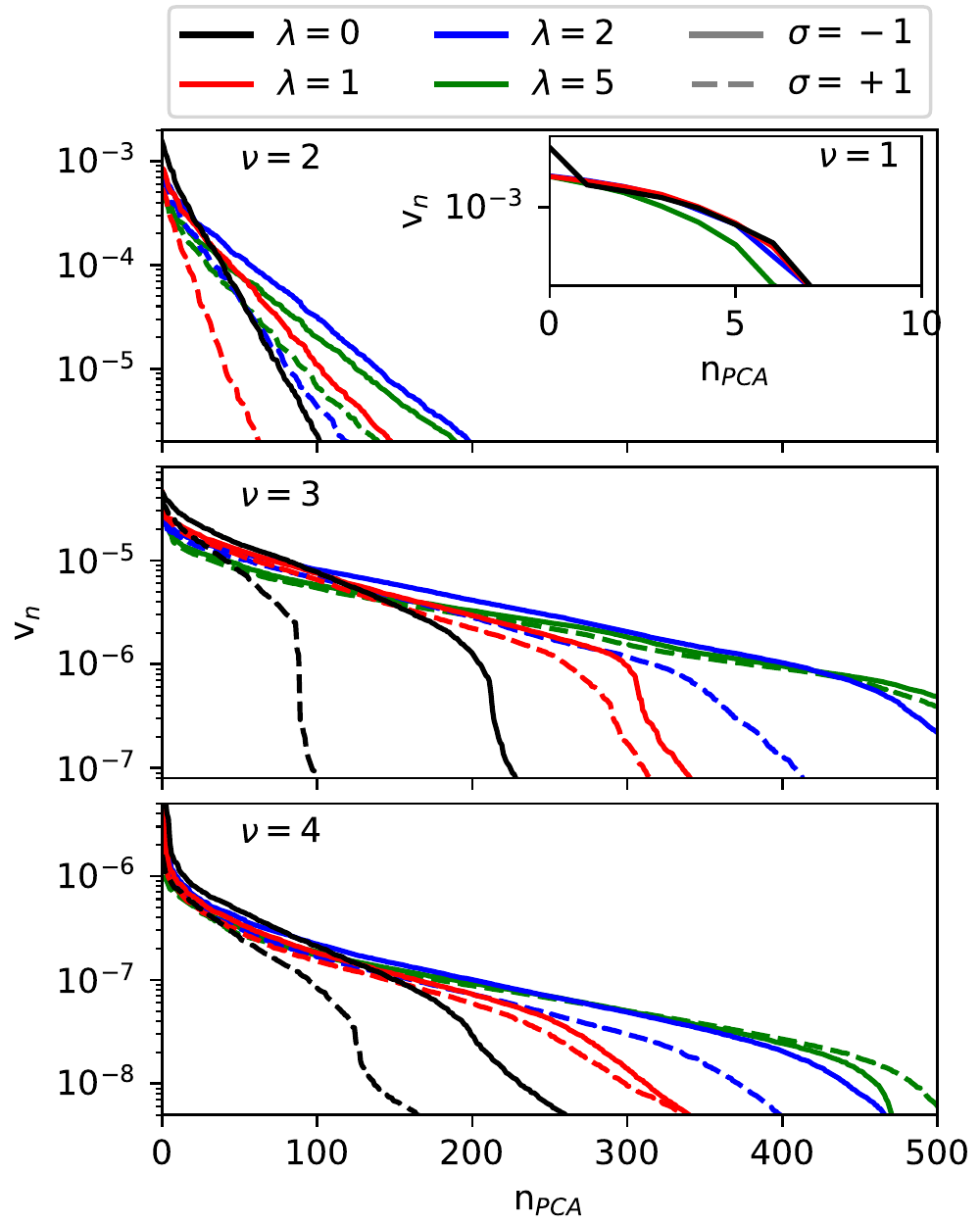}
    \caption{Eigenvalues of the correlation matrix between NICE features of order $\nu$ for 2000 \ce{C}-centred environments extracted from the random \ce{CH4} dataset. Increasing the number of environments does not change significantly the eigenvalue spectra. At each NICE iteration we save 400 most important invariant features, and retain 150 sets of contracted equivariants that are combined with $\ket*{\rho_i}$ in the next iteration.}
    \label{fig:nice-pca}
\end{figure}

\section{Results}

We demonstrate the construction of NICE features on a dataset of 3 million \ce{CH4} quasi-random configurations that has recently been introduced in Ref.~\citenum{pozdnyakov2020completeness}, as the high number of structures and their random nature reveals the behavior of $N$-body invariants in a way that is less biased by the nature of the reference dataset. In a way, this data set represents a worst-case scenario for the iterative contraction, since it does not benefit from a reduction in the intrinsic dimensionality associated with the distribution of input configurations.

\subsection{Correlations between $N$-body features}

We begin by showing how the contraction process behaves during successive body order iterations. We consider \ce{C}-centred features, so each sample in the data set is associated with a single environment. 
In implementing the NICE scheme we apply two additional optimizations. First, we keep track of the eigenvalues $\bv^{\nu;\lambda\sigma}$ associated with the principal components computed at each step, and we use them to ``screen'' the body order iteration~\eqref{eq:contract-recursion}, computing only the terms for which $v^{\nu;ks}_{\tilde{N}} v^{1;l}_{n} $ is greater than a set threshold. 
Second, after each body order iteration and before computing the contraction, we project out the components of the new features that can be expressed as a linear combination of lower order equivariants. 

Fig.~\ref{fig:nice-pca} shows the eigenvalues of $\bC^{\nu;\lambda\sigma}$ at different stages of the procedure. The atom-centred density is expanded on a basis of $n_\text{max}=8$ Gaussian-type radial functions, including angular momentum channels up to $l_\text{max}=6$. Even though we consider a separate density for \ce{C} and \ce{H} atoms, there are only 8 independent $\nu=1$ equivariants, reflecting the fact that the density contribution corresponding to carbon is identical for every \ce{C} centred environment, and thus irrelevant.  
For $\nu=2$ equivariants (corresponding to the most common implementation of $\lambda$-SOAP~\cite{gris+18prl}) the correlation spectrum decays very rapidly, which is consistent with the observation that the power spectrum can be truncated very aggressively with little loss of regression performance -- a fact that has been exploited in several recent scalar and tensorial SOAP-based models~\cite{imba+18jcp,enge+19pccp,gris+19acscs}.
Higher-$\nu$ spectra decay more slowly. Nevertheless, at each body order we considered a few hundreds of features represent 99\%{} of the dataset variance. 
This is in striking constant with the expected exponential scaling of the number of linearly independent equivariants, and underpins the viability of the NICE framework. 
Fig.~\ref{fig:nice-pca} also reflects the importance of contracting separately equivariants of different parity. For $\nu=2$ there is no pseudoscalar component, and all of the pseudotensor features decay faster than the corresponding tensorial equivariant. Even though in this proof-of-principle work we ignore non-asymptotic optimizations, exploiting the different behavior of low-order features of different parities can provide a noticeable reduction in memory and computational requirements.

\begin{figure}
    \centering 
    \includegraphics[width=1.0\linewidth]{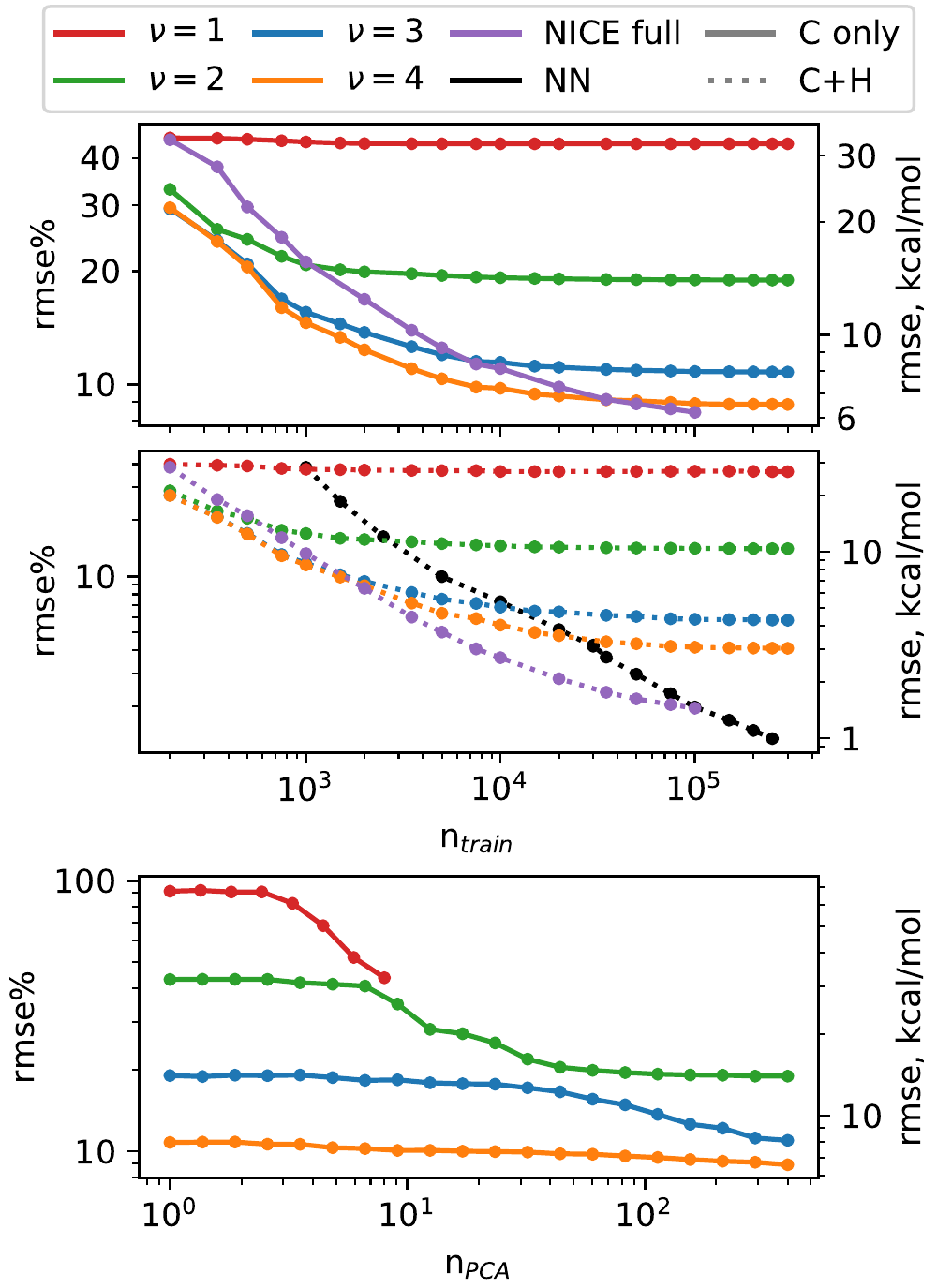}
\caption{Learning curves for the formation energy of \ce{CH4} structures using linear models based on NICE features truncated to increasing body order, a NN model using NICE features up to $\nu=4$, and a $\nu=4$ NICE linear model in which the contraction step was converged fully (using lower $n_\text{max}$ and $l_\text{max}$, see SI). The top panel shows results for models using only \ce{C}-centred features; middle panel uses features centred on both \ce{C} and \ce{H}.
The bottom panel shows the convergence of linear models trained on 100k \ce{CH4} structures, using only \ce{C}-centred NICE features and including increasingly large numbers of PCA components -- i.e. a $\nu=3$, $n_\text{PCA}=100$ model contains all NICE features with $\nu=1,2$ and the top 100 $\nu=3$ features. }
    \label{fig:nice-lc}
\end{figure}

\subsection{Regression performance}

Fig.~\ref{fig:nice-lc} shows learning curves for \emph{linear} NICE models based on increasingly high body-order features. 
In the case of \ce{CH4} structures, \ce{C}-centred features of order $\nu=4$ should in principle provide a complete linear basis to describe the interatomic potential.
In practice, however, the learning curves of linear models saturate at a relatively small train set size. Including terms of increasing $\nu$ delays the saturation, and improves the asymptotic accuracy, but the improvement becomes less dramatic with increasing body order.
The bottom panel of Fig.~\ref{fig:nice-lc} shows that indeed the slower decay of the PCA spectrum is reflected in a slow convergence of the error with the number of PCA components.
While it is possible to systematically improve a linear NICE model by ramping up the number of PCA components (see SI, and the purple curves in Fig.~\ref{fig:nice-lc}), and the number of radial and angular momentum channels, one should contrast this with the use of more flexible models of the target property. 
For instance, a $\approx50$\%{} drop in error can be achieved, by using simultaneously features centred on \ce{C} and \ce{H} atoms. It is clear that, for example, an accurate description of the binding of two \ce{H} atoms in the region far from the carbon is more easily achieved using \ce{H}-centred information.  Furthermore, a neural-network potential based on NICE invariant features avoids saturation altogether, and easily outperforms all linear models in the data-rich limit.  

\section{Conclusions}

The description of atomic structures in terms of features that can be construed as symmetrized $N$-point correlation functions of the atom density has proven to be a very successful approach to construct accurate and transferable machine-learning models of atomic-scale properties. 
A formulation of these representations in terms of a recursion for equivariant features simplifies the calculation of high body-order terms, and can be combined with a contraction step -- which we demonstrate in its simplest form using principal component analysis -- to keep the exponential increase in complexity at bay. 
Even though this $N$-body iterative contraction of equivariants provides a practical approach to construct a systematically-improvable linear basis to model atomic-scale properties, whether doing so constitutes the most robust and computationally-efficient approach to atomistic machine-learning remains an open research problem. 

Systematic benchmarking on more diverse (and less random) data sets, the incorporation of a contraction step informed by supervised-learning criteria~\cite{dejo-kier92cils,kpcovr-arxiv}, as well as the use of sparse feature selection methods~\cite{imba+18jcp,li2018-lassoselection}, the combination with $N$-point correlation features designed to treat long-range physics~\cite{gris-ceri19jcp},
and ultimately the comparison with kernel methods and more general non-linear regression strategies~\cite{anderson2019cormorant}  
are just some of the many lines of investigation that can be pursued based on the NICE framework to increase the accuracy and reduce the computational effort of atomistic machine learning. 

\section*{Acknowledgments}
MC and SP acknowledge support from the Swiss National Science Foundation (Project No. 200021-182057). JN was supported by a MARVEL INSPIRE Potentials Master's Fellowship. MARVEL is a National Center of Competence in Research funded by the Swiss National Science Foundation.

\end{document}